# Sequence Dependence of Self-Interacting Random Chains

*Anders Irbäck*[1] and *Holm Schwarze*[2]

Department of Theoretical Physics, University of Lund
Sölvegatan 14A, S-223 62 Lund, Sweden

## Abstract

We study the thermodynamic behavior of the random chain model proposed by Iori, Marinari and Parisi, and how this depends on the actual sequence of interactions along the chain. The properties of randomly chosen sequences are compared to those of designed ones, obtained through a simulated annealing procedure in sequence space. We show that the transition to the folded phase takes place at a smaller strength of the quenched disorder for designed sequences. As a result, folding can be relatively fast for these sequences.

[1]Internet: irback@thep.lu.se .

[2]Internet: holm@thep.lu.se , new address: NEuroTech A/S, Fruebjergvej 3, 2100 Copenhagen, Denmark.

# 1 Introduction

The ability of natural proteins to fold into globular states of well-defined shape is a fascinating feature, which has attracted considerable attention (see e.g. [1, 2]). However, it is still not understood, how the information about the three-dimensional *native* structure of a protein is coded in the linear sequence of its constituencies and how proteins search the configuration space on reasonable timescales to find the native state [3].

In recent years there has been an increasing interest in understanding the essential mechanisms that govern the folding of real proteins by studying simplified models of random self-interacting chains or random heteropolymers [4–8]. Although these models do not share the rich details of biological proteins there is hope that a better understanding of such models will help to find and understand the universal features responsible for the unique behavior of real proteins. An analytic treatment of these models has been put forward within mean field theory employing the replica method [5, 6, 9], but the problem is yet far from solved [10].

Recently, Iori, Marinari and Parisi (IMP) [7] have suggested a model of a random self-interacting chain displaying features, that make it interesting for the understanding of real proteins. In particular, IMP have given evidence for the existence of a folded glassy phase with long relaxation times and few stable states of fairly well-defined shapes.

Following their approach, we will be concerned with a chain of $N$ monomers in a continuous three-dimensional space. The interaction between the monomers is modeled by a Lennard-Jones potential whose attractive part contains a quenched disorder. We study the thermodynamic properties of this model using the Hybrid Monte Carlo method [11] which has recently been applied successfully to the simulation of polymers with self-repulsion [12]. For randomly chosen realizations of the quenched disorder we obtain results similar to those found by IMP. These results confirm that the generic behavior, at strong quenched disorder, exhibits interesting similarities to that of globular proteins. For the typical chain, however, it appears impossible to identify a state corresponding to the unique native structure of a protein. Therefore it is of interest to find out whether there are special sequences of interactions along the chain for which there exists a state that is thermodynamically dominant, and kinetically accessible in reasonable times. Shakhnovich and Gutin [13, 14] have proposed a thermodynamically oriented method for finding such sequences and have applied it to a lattice model for protein folding. Using the present model, we test the generality of this method for sequence design, which involves a simulated annealing procedure in sequence space. By varying the strength $\epsilon$ of the quenched disorder, we study the transition to the folded phase and the behavior in this phase for designed and random sequences. The results obtained for these two kinds of sequences differ markedly, even though some quantities, like the radius of gyration, show a fairly weak sequence dependence. The most important difference is that the dominance of a single state sets in at a relatively small $\epsilon = \epsilon_{\mathrm{F}}$ for designed sequences. As the dynamics tends to be faster when the quenched disorder is weaker, this means that indeed the designed sequences are more likely to satisfy both the thermodynamic and the kinetic requirements for folding. Notice that the behavior of sequences with small $\epsilon_{\mathrm{F}}$ should be similar to that of sequences with high folding temperature. In fact, using a lattice model with contact interactions, Šali *et al.* have shown that there is strong correlation between high folding temperature and good folding properties [15].

The paper is organized as follows: In Sec. 2 we describe the model introduced by IMP and the Hybrid Monte Carlo method used in our simulations. In Sec. 3 we discuss some results about the structure of the energy landscape of the random self-interacting chain that can be extracted from our simulations. In Sec. 4 we discuss the folding transition of the model and investigate the influence of the sequence of interactions on this behavior. Finally we conclude with a summary.

## 2 The Model

Throughout this paper we consider a linear chain of $N$ sites or monomers, whose positions in three-dimensional continuous space are given by $\underline{x}_i$, $i = 1, \ldots, N$. We eliminate the overall translational degree of freedom by assuming that the center of mass is held fixed at the origin, i.e. $\sum_{i=1}^{N} \underline{x}_i = 0$. Alternatively, the system can be described by the $N-1$ bond vectors $\underline{b}_j = \underline{x}_{j+1} - \underline{x}_j$, $j = 1, \ldots, N-1$, connecting adjacent sites. The Hamiltonian of the system is given by

$$H = \sum_{i<j} \left( \delta_{i,j-1}\, r_{ij}^2 + \frac{R}{r_{ij}^{12}} - \frac{A + \epsilon \eta_{ij}}{r_{ij}^6} \right) \qquad (1)$$

where $r_{ij} = ||\underline{x}_i - \underline{x}_j||$ is the Euclidean distance between sites $i$ and $j$. The first term in Eq. 1 is just a harmonic attraction which holds adjacent sites together and enforces the chain structure. The remaining terms are a standard Lennard-Jones (LJ) potential, whose van der Waals attraction contains a quenched disorder. The $\eta_{ij}$'s are independent random variables drawn from a uniform distribution with zero mean and unit variance. Hence, the potential between sites $i$ and $j > i+1$ has a minimum at $r_{ij} = [2R/(A + \epsilon\eta_{ij})]^{1/6}$ if $\eta_{ij} > -A/\epsilon$. This model has been studied by IMP using standard Monte Carlo techniques in three dimensions and subsequently by Fukugita *et al.* [8] in two dimensions. IMP studied the behavior of the model for different values of $A$ and $\epsilon$, keeping $R = 2$ and $k_B T = 1$ fixed. For $\epsilon = 0$, they showed that a transition occurs near $A = 2$, from a coil phase at small $A$ to a shapeless globule phase. Using $A = 3.8$, they then found glassy behavior with few stable states for sufficiently large strength $\epsilon$ of the disorder. These states persisted for long times in the Monte Carlo simulations, and the system came back to the same state after having visited a few completely different ones. The transition from the shapeless globular phase to the glassy phase was found to be very abrupt. Fukugita *et al.* studied the same model in two dimensions and explored the level structure of the lowest-lying states. Their results show the emergence of a single global minimum with a wide gap to the next lowest energy level as $\epsilon$ increases. In addition to the $(A, \epsilon)$ phase structure, Fukugita *et al.* investigated the temperature phase structure of the model. For a folding sequence, they found that the transition from coil state at high temperature to unique folded state at low temperature takes place through a shapeless globule phase. This behavior has also been observed using a different model [16].

To simulate the behavior at constant temperature of the system defined by the Hamiltonian in Eq. 1 we have employed the Hybrid Monte Carlo method [11]. A discussion of the use of this general method in simulating proteins can be found in Ref. [17], while the implementation used here is described in Ref. [12]. We give a brief description of this scheme, and refer to Ref. [12] for further details. The simulation is described most easily by using the bond variables $\underline{b}_i$. It is based on the evolution arising from the fictitious Hamiltonian

$$H_{\mathrm{MC}} = \frac{1}{2}\sum_{i=1}^{N-1} \underline{\pi}_i^2 + \frac{1}{k_{\mathrm{B}}T} H(\underline{b}_1, \ldots, \underline{b}_{N-1}) \tag{2}$$

where $\underline{\pi}_i$ is an auxiliary momentum variable conjugate to $\underline{b}_i$. The first step in the algorithm is to generate a new set of momenta $\underline{\pi}_i$ from the equilibrium distribution $P(\underline{\pi}_i) \propto \exp(-\frac{1}{2}\underline{\pi}_i^2)$. Starting from these momenta and the old configuration, the system is evolved through a finite-step approximation of the equations of motion. The configuration generated in such a trajectory is finally subject to an accept-or-reject question. The probability of acceptance in this global Metropolis step is $\min(1, \exp(-\Delta H_{\mathrm{MC}}))$, where $\Delta H_{\mathrm{MC}}$ is the energy change in the trajectory. This accept-or-reject step removes errors due to the discretization of the equations of motion. The algorithm has two parameters, namely the step size $\delta$ and the number of steps in each trajectory, $n$. We have used a fixed trajectory length, $n\delta = 1$, and the remaining free parameter has been adjusted so as to have a reasonable acceptance rate.

## 3 The Structure of the Energy Surface

Using the Hybrid Monte Carlo method, we have simulated the behavior at constant temperature of the model described in the previous section. To extract more information about the energy landscape from these simulations, we have employed a deterministic quenching procedure which brings the system to a nearby local energy minimum. This minimization has been carried out by using a conjugate gradient method. The quenching procedure provides information about the energy level spectrum, and has been applied to models for protein folding in Refs. [18, 8]. The removal of thermal noise is very useful when studying the evolution of the system.

In Fig. 1 we show the evolution of the actual and the quenched energy in a Monte Carlo run for $N = 8$. Following IMP, we have here used $A = 3.8$, $R = 2$, $\epsilon = 6$ and $k_{\mathrm{B}}T = 1$. Throughout this paper we will only vary the values of $A$ and $\epsilon$, i.e. the relative strength of the deterministic and random parts of the attractive contribution to the LJ-potential, while the values of $R$ and $k_{\mathrm{B}}T$ will be kept fixed. IMP have reported runs which indicate that a change in $T$ roughly corresponds to rescaling of the other parameters. While the unquenched data are very noisy, the quenched measurements show a clear structure. The system visits only a small number of different energy levels and spends more than 99% of the time in the lowest eight of these. From the point of view of dynamics, these eight levels can be divided into two groups, each consisting of four levels. The system moves relatively easily between levels belonging to the same group, while transitions from one group to the other are rare; only less than 1% of all transitions are between states of different groups. Therefore, these groups correspond effectively to two different thermodynamic states, separated by a high free-energy barrier.

A study of the difference between the configurations corresponding to these levels suggests that each group can be thought of as a fairly narrow "valley" in the free-energy landscape. To get a measure of the distance between two configurations with distance matrices $r_{ij}^a$ and $r_{ij}^b$, it is convenient to use a quantity introduced by IMP which employs

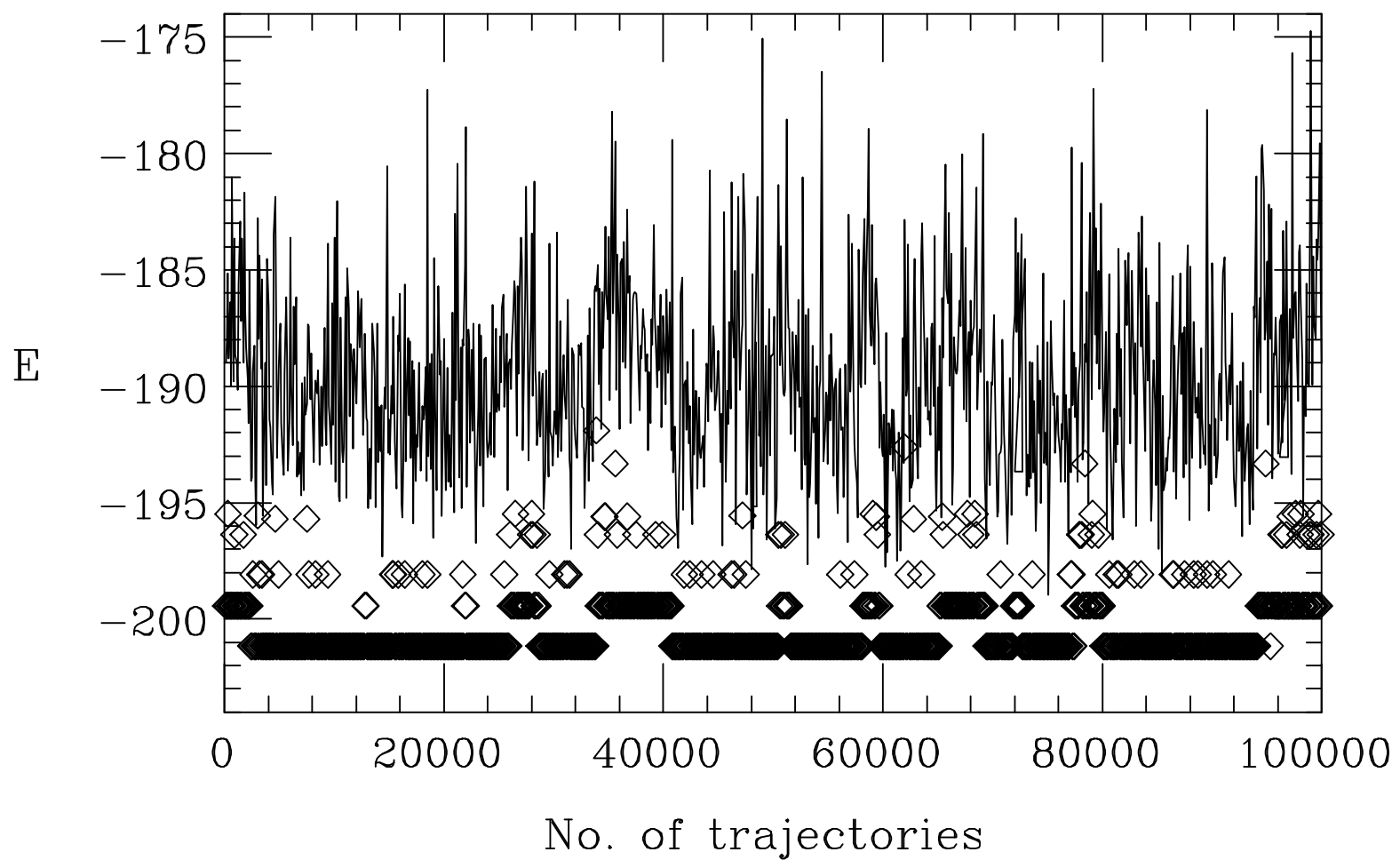

Figure 1: Monte Carlo evolution of the energy (the data points are connected by a line) and the quenched energy (diamonds) for $N = 8$, $A = 3.8$ and $\epsilon = 6$. Measurements have been taken every 100 iterations.

the LJ part of the energy between couples of sites, given by

$$\Delta^2_{ab} = \frac{1}{N(N-1)} \sum_{i>j} (E^a_{ij} - E^b_{ij})^2 \tag{3}$$

where

$$E^a_{ij} = \frac{R}{r^{a\,12}_{ij}} - \frac{A}{r^{a\,6}_{ij}} \tag{4}$$

and similarly for $E^b_{ij}$. A closely related measure, for monitoring dynamical evolution, has been introduced by Thirumalai *et al.* [19] and applied to protein folding in Ref. [20]. The parameters $A$ and $R$ in Eq. 4 are chosen as in the corresponding Monte Carlo run, although the results are not sensitive to this choice. This definition does not require the elimination of any translational, rotational or reflectional degrees of freedom and emphasizes local differences in the energetic state of the sites. The distances between the configurations corresponding to the eight energy levels discussed above are indeed compatible with the existence of two groups of structurally and thus energetically similar states. The distances between states within the same group all lie in the interval $0.07 < \Delta^2_{ab} < 0.17$ and are clearly smaller than those between states of different groups which all lie in the interval $0.5 < \Delta^2_{ab} < 0.8$.

It is important to note that these results are for one particular realization of the disorder. While we found other chains with a similar splitting of the energy levels into different groups, it is for $N = 8$ easy to obtain other examples of realizations of the disorder for which distinct groups do not appear.

The existence of different valleys in the free-energy landscape affects the dynamics of the system even more strongly as the system size is increased. In Fig. 2 we show results of five different simulations for a chain with $N = 16$. The values of $A$, $R$, $\epsilon$ and $k_{\rm B}T$ are the same as before. The difference between these runs is that they were started from different configurations, while the realization of the disorder was not changed. After an initial period of relaxation, the system moves only between a small number of different levels. The sets of levels visited in the different runs are disjoint, which shows that there are several different groups of states, and in each run only one or very few of the corresponding valleys are explored. This result is in agreement with the assumption that the heights of the barriers separating different valleys — and thus the time necessary to cross them — increases with the system size.

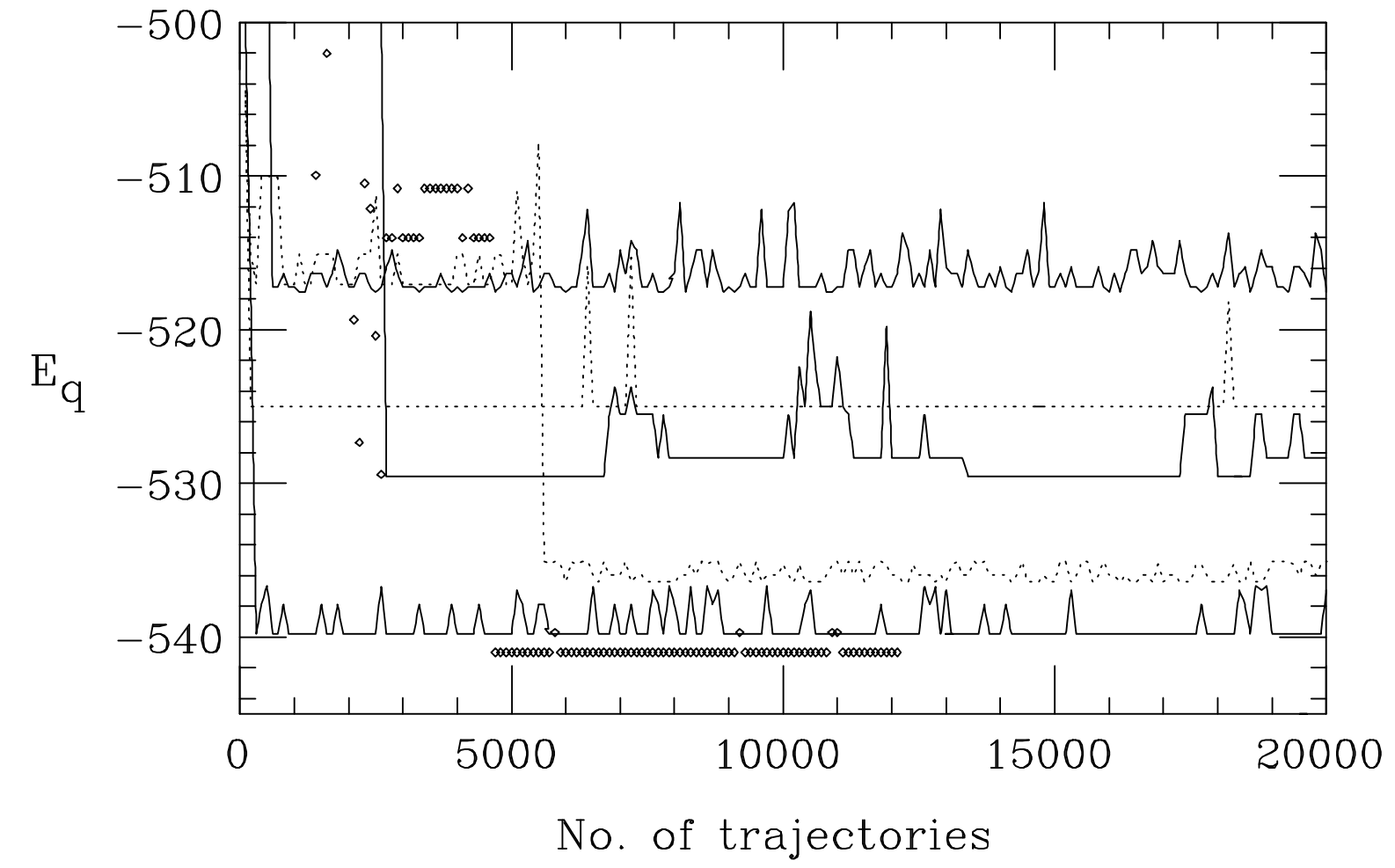

Figure 2: Monte Carlo evolution of the quenched energy for a chain with $N = 16$, $A = 3.8$ and $\epsilon = 6$. The full and dotted lines are results obtained using ordinary Monte Carlo and five different initial configurations. The diamonds are the result of a run in which the auxiliary Hamiltonian has been used, as explained in the text. Measurements have been taken every 100 trajectories.

The structure of the free-energy landscape makes a numerical study of these systems computationally very demanding. Therefore, it would be desirable to develop improved Monte Carlo schemes in order to efficiently explore the configuration space. We have tested a simple strategy with a more modest aim, namely to find the low-lying energy levels faster. To motivate this scheme, we first note that the monomers act to some extent as hard spheres, due to the repulsive $r^{-12}$ term. Minimization of the energy involves finding out how to pack these spheres under the influence of the attractive terms. In general, this is, of course, a very difficult task, but a simplification may occur for large $\epsilon$. If $\epsilon$ is large then the depths of the LJ potentials vary considerably, between zero (purely repulsive potential) and $(A + \epsilon\sqrt{3})^2/4R$. It may then be advantageous to pay particular attention to the strongest interaction terms, at least if the chain is not very long. The procedure considered here is a simple way to do that. As before we use the Hybrid Monte Carlo algorithm, but now we let the evolution of the system be governed by an auxiliary Hamiltonian. Initially we take this Hamiltonian to be the full Hamiltonian with all $r^{-6}$ terms removed. In the absence of these terms, the system is not frustrated and therefore easy to simulate. To this Hamiltonian we first add the $r^{-6}$ term with largest $\eta_{ij}$, then the term with next largest $\eta_{ij}$, and so on. This is continued until all the $r^{-6}$ terms have been included. After each change of the auxiliary Hamiltonian, a short simulation is performed. In our calculations each of these simulations consisted of 100 trajectories and the trajectory length was the same as before. We have tested this procedure on the chain studied in Fig. 2, using again five different initializations. In the present case, the results from the various runs were similar. In particular, the sets of energy levels visited at the end of the runs were identical. The evolution of the quenched energy in one of these runs is given in Fig. 2. This figure shows that the procedure is successful in the sense that new energy levels show up which are lower than all those found in the ordinary Monte Carlo runs. We stress that the original Hamiltonian and not the auxiliary one has been used in these measurements. Therefore, we find it remarkable that these low-lying levels appear for the first time already when less than half of the $r^{-6}$ terms are present. The weakest 60% of these terms seem to influence the structure of the chain very little. The procedure described here will be discussed in more detail in a forthcoming paper [21].

Even though there are several differences, it may be of interest to compare the potential considered here with a more realistic one. For this purpose we consider the simplified potential for structure determination that was proposed in Ref. [22]. A central ingredient of this potential is a species-dependent LJ interaction, which represents the effects of hydrophilicity and hydrophobicity. Each amino-acid residue is described by one backbone site and one or two sites representing the side chain. To make a comparison possible, we have therefore converted the LJ parameters used in [22] into parameters appropriate for the chain representation used here. Using these results, we have generated distributions of the LJ parameters by weighting the residues according to their frequency of occurence in proteins [2]. The distribution of the $r^{-6}$ coefficient allows both positive and negative values. Averaging over potentials with attractive $r^{-6}$ term, we find that the mean and variance of the depth of the potential, $-H_{min}/k_{\mathrm{B}}T$, are 3.1 and 4.5, respectively. In the model considered in this paper, the corresponding values are 8.4 and 56.3 for the parameters studied above. In order to obtain a distribution of potentials that is more closely related to the one used in Ref. [22] we will use the parameters $A = 0$ and $R = 2$ in the following. For this choice and for $\epsilon = 4.5$ the corresponding mean and variance are 2.5 and 5.1. Furthermore, due to the absence of extremely deep minima the dynamics can be expected to be somewhat faster for this choice of parameters.

## 4 Optimizing the Realization of the Disorder

In the previous section we have seen the strong dependence of the dynamical behavior of this model on the actual choice of the random part of the interaction. This observation leads to the question of how the behavior of an "optimized" or "selected" sequence would differ from that of randomly chosen chains. In particular we will be concerned with the folding properties of the IMP model and its dependence on the realization of the quenched interactions. To do this we follow an idea proposed by Shakhnovich *et al.* [13, 14] of using "protein design" to study the folding properties of model proteins with selected sequences. Shakhnovich *et al.* studied chains of length $N = 80$ on a cubic lattice with a contact interaction between them. In order to find a sequence that provides a very low energy on an arbitrarily chosen target structure, they used a simulated annealing procedure in sequence space keeping the positions $\underline{x}_i$ of the sites fixed. In subsequent lattice MC runs in configuration space with the optimized sequence kept fixed the designed chains folded into the unique target configuration after about $10^6$ MC steps. The energy spectra of the designed chains were characterized by a global energy minimum separated from the remaining energy levels by a pronounced gap. It was argued that the existence of this energy gap in the designed chains is responsible for the fast folding times, because out of a large number of randomly generated sequences only those which showed such a gap in their energy spectrum folded fast [14, 15]. Furthermore, it is interesting to note the relation between the thermodynamic design of the chains and the problem of kinetic accessibility of the target configuration. The sequences that minimized the energy of a given target configuration were also able to find this structure during the subsequent MC simulation in reasonable times [14].

In order to study the sequence dependence of the properties of the IMP model we have concentrated on a specific choice of the parameters for the deterministic part of the LJ potential ($A = 0, R = 2$). Within the range of $\epsilon$ values considered here, this choice produces distributions of LJ potentials which seem reasonable compared to more realistic models as discussed in the previous section. The results presented below have been obtained for $N = 16$ and three different realizations of the $\eta_{ij}$'s, which will be referred to as sequence 1,2, and 3, respectively.

As the relative strength $\epsilon$ of the random part of the potential is increased we observe a behavior which is in agreement with the one reported in the original work by IMP [7]: For $\epsilon = 0$ the system is in an open coil state characterized by a large end-to-end distance $r_{\rm ee} =< ||\underline{x}_1 - \underline{x}_N|| >$ and a large radius of gyration $r_{\rm gyr} = \sqrt{< \frac{1}{N} \sum_{i=1}^{N} \underline{x}_i^2 >}$. The system can move around rather freely in the energy landscape and visits a large number of local minima (see Fig. 3).

At around $\epsilon \approx 3.5$ we find a transition to a globule phase indicated by a drop in $r_{\rm ee}$ and $r_{\rm gyr}$ (see Fig. 4). In this phase the chain has compactified, but the individual monomers can still move around freely. Correspondingly, the system still visits a large number of states as can be seen from Fig. 3.

As mentioned in Sec. 2, IMP showed that the system is in a glassy phase for large $\epsilon$ if $A = 3.8$. Our results show that the large-$\epsilon$ behavior is glassy also for $A = 0$. For $A = 0$, the transition to the glassy phase takes place near $\epsilon = 6.5$. The transition is not observable in the behavior of $r_{\rm ee}$ and $r_{\rm gyr}$ (see Fig. 4) but can clearly be detected by a dramatic increase in the autocorrelation times. This is illustrated in Fig. 5, where we show estimates of the integrated autocorrelation time of the quenched energy, $\tau_{\rm E_q}$. The integrated autocorrelation time $\tau_{\rm O}$ of a quantity $O$ is given by $\tau_{\rm O} = \frac{1}{2} \sum_{t=-\infty}^{\infty} C(t)/C(0)$, where $C(t) = \langle O(t)O(0) \rangle - \langle O \rangle^2$ is the correlation between measurements separated by $t$ trajectories. The fact that $\tau_{\rm E_q}$ is much larger at $\epsilon = 7$ than at small $\epsilon$ can also be seen directly from Fig. 3. At $\epsilon = 7$, the system tends to move between a small number of energy levels over extended periods time and revisits the same set of levels after several hundreds of trajectories. In Fig. 6 the relative number of times $\delta$ the system was found in the most frequently visited quenched-energy state is plotted against $\epsilon$. While $\delta$ is below 10% for small values of $\epsilon$ there is an increase in the dominance of a single state if $\epsilon$ is increased above 6.5.

In the following we investigate how this behavior is altered if the entries of the interaction matrix $\eta_{ij}$ are not placed randomly along the chain but such that the energy for a certain target configuration is minimized. An arbitrary target configuration (see Fig. 7) was created randomly with a radius of gyration of $r_{gyr} \approx 1.1$, a typical value for the folded chains with randomly placed interactions. For a given realization $\eta_{ij}$ of the random part of the interaction matrix we have permuted the 120 entries as to minimize the energy of Eq. 1 for the target configuration. In order to search the approximately $10^{200}$ possible permutations we have used straightforward simulated annealing in sequence space. This procedure ensures that the statistics of the $\eta_{ij}$'s remains unchanged in the optimized sequence, since only the indices are permuted without changing the values of the interactions. Starting from the three random sequences studied above, we have created their optimized counterparts (called sequences $1^*$, $2^*$ and $3^*$, respectively) which yielded the lowest energies found for the target configuration. Subsequently, we have performed Hybrid Monte Carlo simulations in configuration space for the sequences starting from random coil-like initial configurations. The results are shown in Figs. 3, 4 and 6.

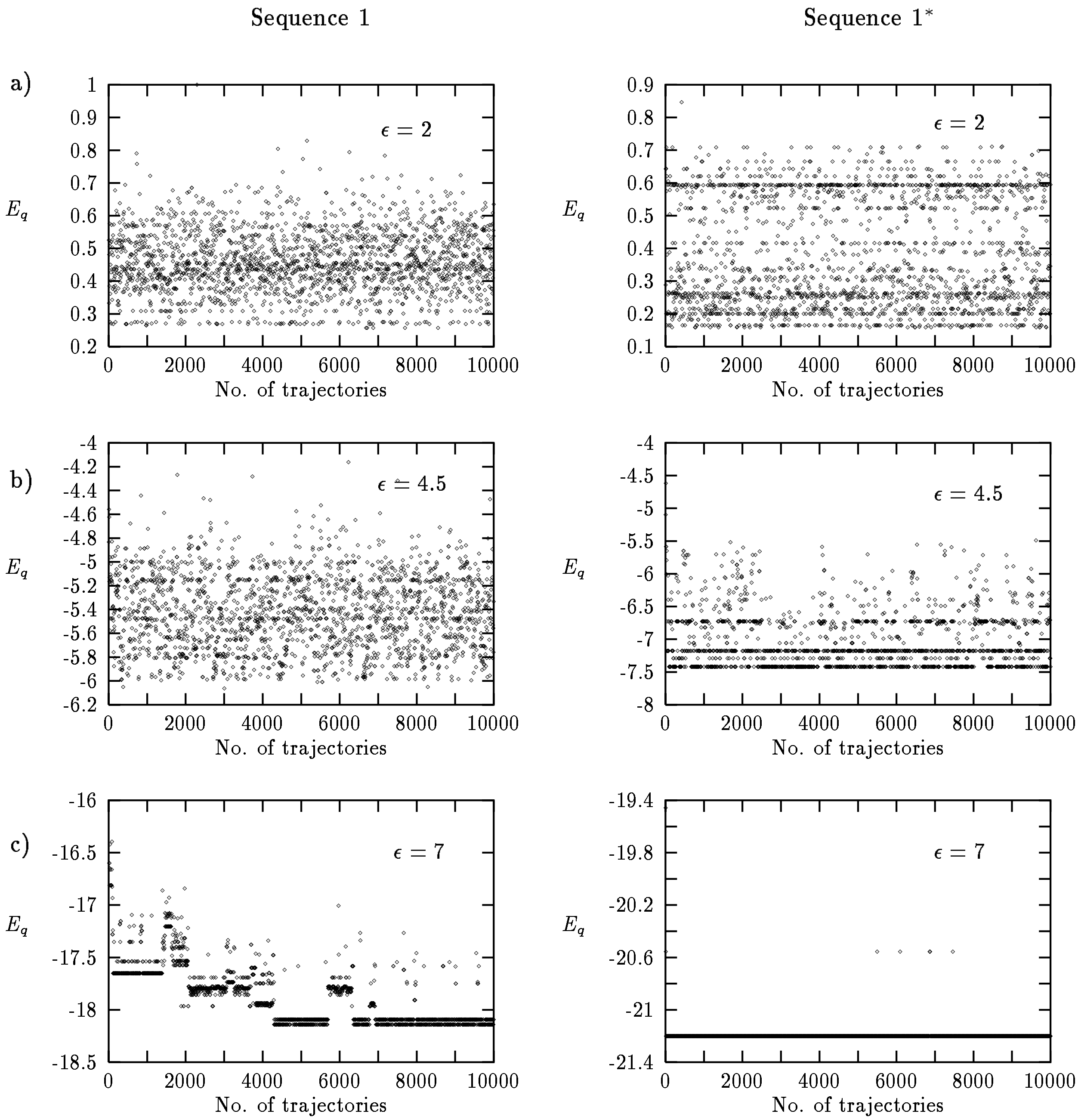

Figure 3: Evolution of the quenched energy for $N = 16$, $A = 0$ and different values of $\epsilon$, sampled every 5 trajectories: a) $\epsilon = 2$, b) $\epsilon = 4.5$, c) $\epsilon = 7$. Left column: randomly chosen sequence (sequence 1). Right column: same set of $\eta_{ij}$'s, but after optimization (sequence 1*).

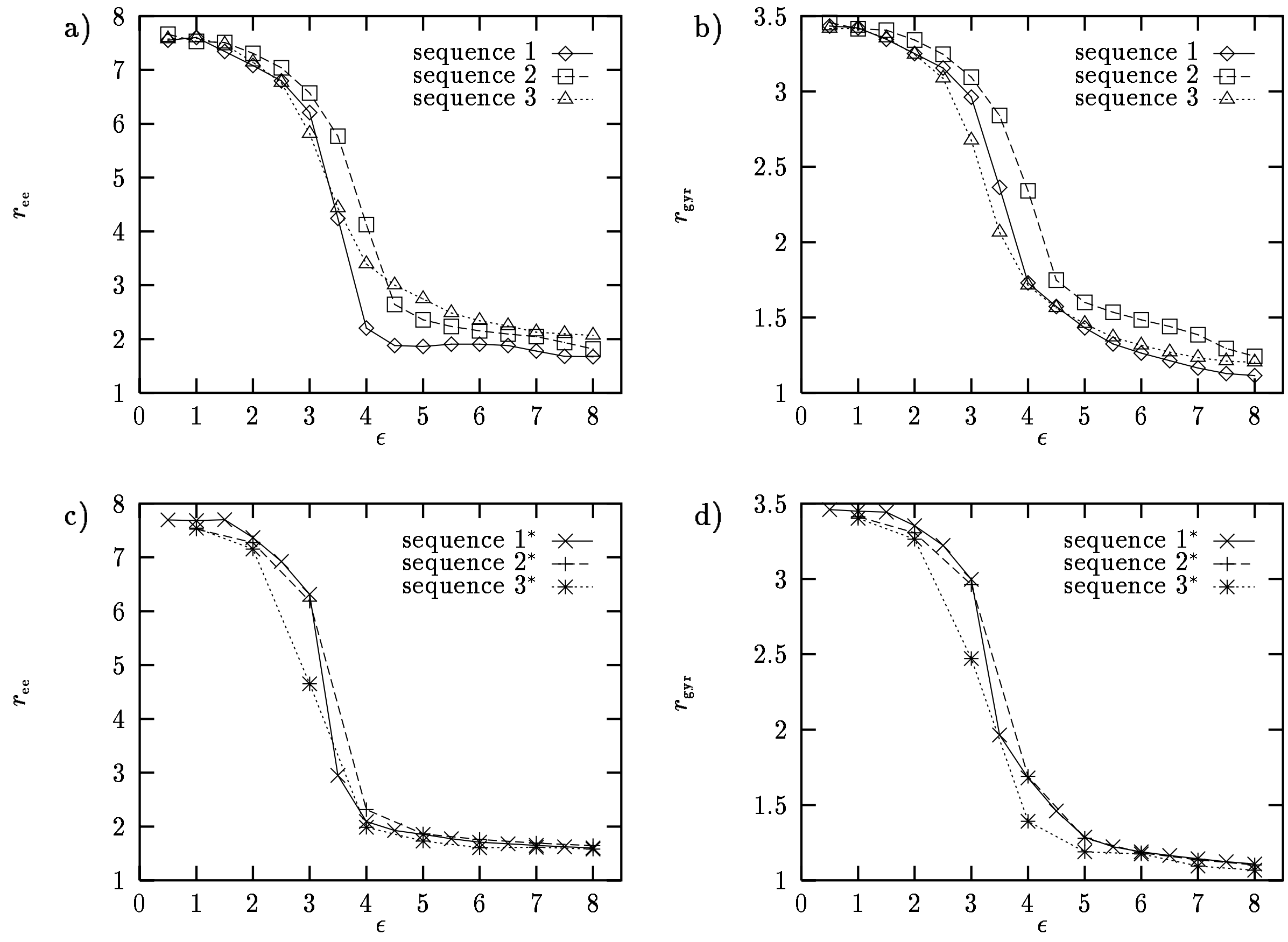

Figure 4: End-to-end distance (left) and radius of gyration (right) as functions of $\epsilon$ for different random sequences (top) and the respective optimized sequences (bottom).

As can be seen from Fig. 4, there is no significant sequence-dependence detectable in the macroscopic quantities $r_{ee}$ and $r_{gyr}$. However, the structure of the free energy of the optimized sequences differs significantly from that of the random chains, as reflected by the dynamical evolution. The dominance of a single quenched-energy state sets in at much smaller values of $\epsilon$ than for random sequences (see Figs. 3 and 6). The evolution of quenched energies of sequence $1^*$ in Fig. 3 shows that already for small values of $\epsilon$ a few energy levels are more favored than others. As $\epsilon$ is increased the formation of a small number of dominant low-lying states already occurs at around $\epsilon \approx 4$, almost simultaneously to the compactification at $\epsilon \approx 3.5$. It is remarkable that the dominant states are already found after a few hundred trajectories and typically dominate the dynamical evolution over the entire run. At $\epsilon = 7$, chain $1^*$ gets trapped after a few trajectories in a single state, which is identical to the quenched state that would be reached if the quenching was started from the target configuration.

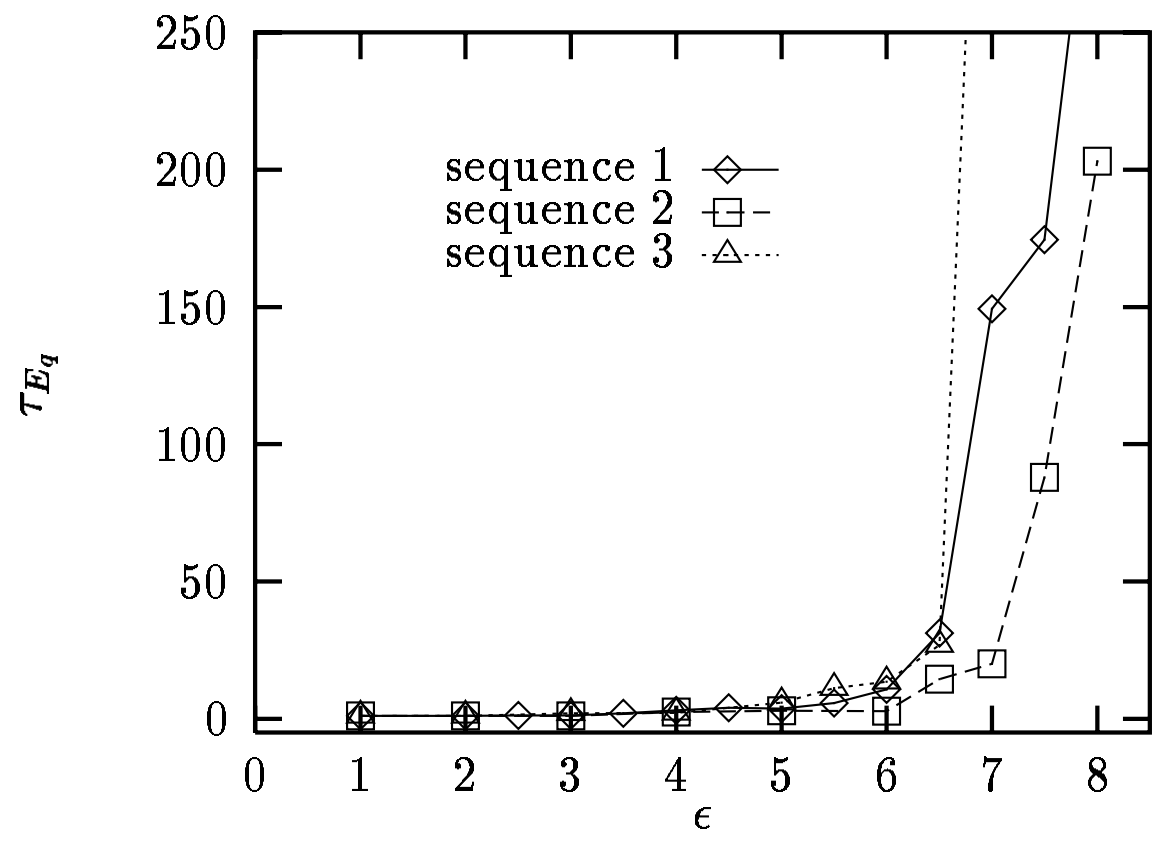

Figure 5: Autocorrelation times of the quenched energies for the three random sequences.

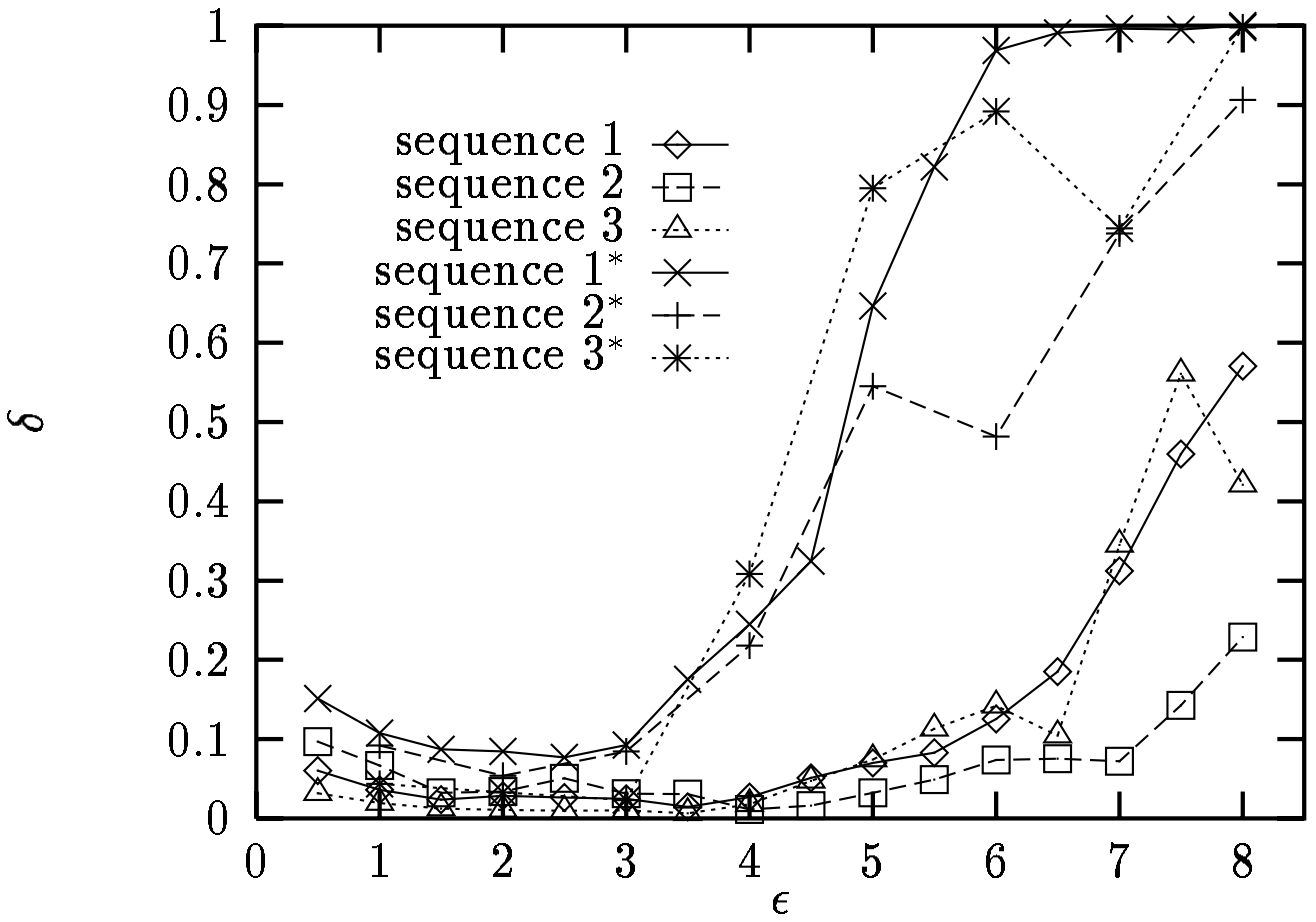

Figure 6: The time spent in the most frequently visited quenched-energy state relative to the others. Results are shown for the three random sequences and the respective optimized sequences.

The results for the sequences $2^*$ and $3^*$ are qualitatively similar. However, in those cases we found two dominating states in the respective large-$\epsilon$ regimes, and the systems freqently moved between them. This behavior is reflected by the somewhat lower $\delta$ values in Fig. 6. Furthermore, at $\epsilon = 8$ sequence $3^*$ did not find the target configuration immediately as it was the case for the first two sequences. Chain 3* spent the first 10,000 trajectories above a different local minimum with somewhat higher energy before the transition to the target configuration occured. Then, the target state dominated for the remaining 90,000 trajectories of this run. However, for $\epsilon$ values just above 4.5 and up to $\epsilon \approx 7$, sequence 3* displayed a fast dynamics as observed for the other chains.

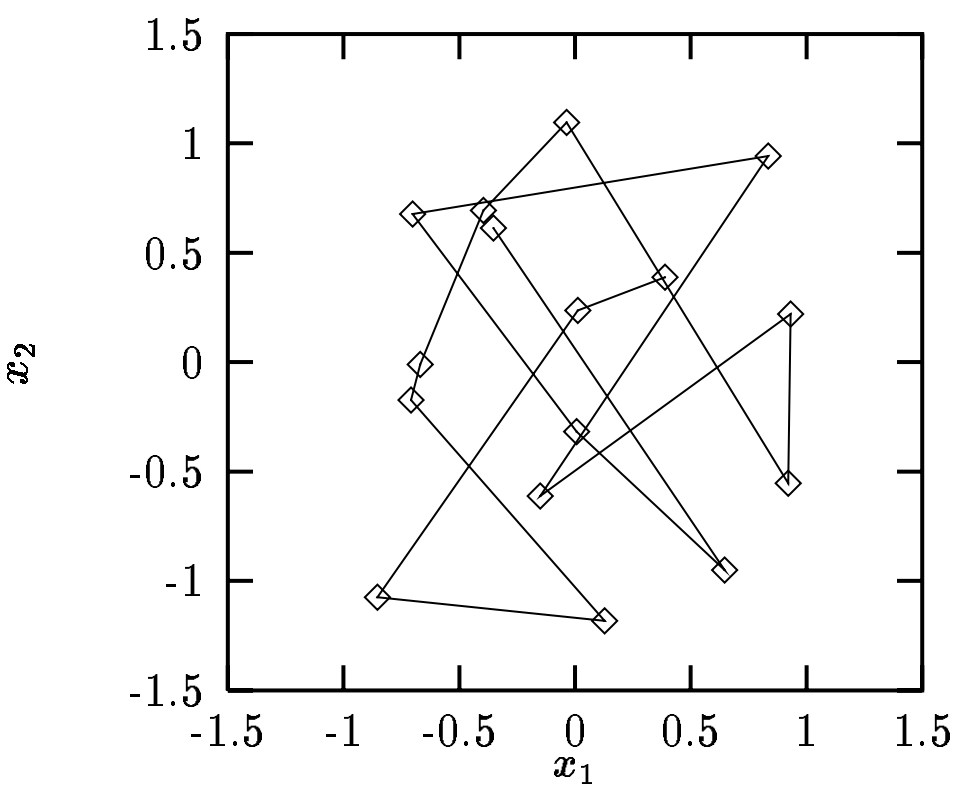

Figure 7: Projection of the target configuration used for the optimization of the sequence of interactions.

Let us finally summarize the results for the folding properties of optimized and random sequences. The main question addressed is whether the different sequences exhibit a thermodynamically dominant state which, under the same conditions, is kinetically accessible. From Fig. 6 it can be seen that a significant weight of a single state is obtained already at relatively small $\epsilon$ for optimized sequences. In fact, $\delta$ is roughly the same for random sequences at $\epsilon = 7$ as for optimized ones at $\epsilon = 4.5$. This difference in $\epsilon$ has a strong effect on the dynamics of the two systems, as can be seen from Fig. 3. While the evolution of $E_q$ for the random sequence at $\epsilon = 7$ is dominated by different, well-separated *groups* of levels, it shows frequent transitions between all states involved for the optimized sequence at $\epsilon = 4.5$. This observation indicates the presence of high energy barriers in the random case at $\epsilon = 7$, which slows down the dynamics considerably. In the optimized case, on the other hand, high energy barriers seem to be absent. Therefore, folding times should be much shorter. From these results we conclude that optimized sequences are much more likely to satisfy both the thermodynamic and the kinetic requirements for folding. The design method of Refs. [13, 14] is therefore applicable also to the present model.

# 5 Summary and Outlook

In this paper we have studied the random heteropolymer model proposed by IMP. Our main focus has been on the question of how the thermodynamic behavior depends on the sequence of interactions along the chain. This has been studied for a set of parameters slightly different from that used by IMP. The choice considered here gives rise to similar thermodynamic behavior and has been made to get a somewhat faster Monte Carlo evolution. Our comparison with the more realistic model of Ref. [22] suggests that the parameters chosen are reasonable.

For the choice of parameters originally used by IMP we have found a strong dependence of the dynamical behavior on the specific choice of the disorder. We have demonstrated that for some chains there exist different valleys in the free-energy surface, each of which contains a number of local minima. As the system size is increased it gets more and more difficult to cross the free-energy barriers between these valleys, and the system tends to explore only one or a few valleys. We have found that the behavior is similar for $R = 2$, $A = 0$ and strong quenched disorder. The results mentioned so far refer to randomly chosen sequences. In addition to those, we have studied designed sequences, obtained by the thermodynamically oriented procedure suggested in Ref. [13]. The thermodynamic behavior of these two kinds of sequences has been compared. The results show that the extent of the chain, as measured by the radius of gyration, depends only weakly on the sequence. In particular, they show that the compactification of the chain takes place at approximately the same strength of the quenched disorder, $\epsilon$, for different sequences. At large $\epsilon$, the density of low-lying energy levels is low and the bottom part of the spectrum dominates thermodynamically. The sparseness of the bottom part of the spectrum sets in earlier for designed sequences. For these sequences it is possible to find values of $\epsilon$ where the lowest level has a large statistical weight and can be reached fast. In order to have a significant weight of a single state for random sequences, it is necessary to use a considerably larger $\epsilon$ which makes the folding process much slower. Hence, the thermodynamic design strategy suggested by Shakhnovich *et al.* is also applicable here and yields chains that tend to fold faster than there random counterparts.

Even though the IMP model displays interesting features for comparably short chains, it would of course be desirable to study larger systems, in order to investigate the size dependence of our results. However, due to the slow dynamics in the glassy phase this demands very long simulations or more efficient Monte Carlo schemes. We have demonstrated a simple strategy to find the low-lying states for strong quenched disorder, but to find a suitable algorithm for the exploration of the thermodynamic properties of long random heteropolymers remains a challenging problem.

**Acknowledgments**
We are grateful for useful discussions with Bo Jönsson, Giorgio Parisi, Carsten Peterson and Magnus Ullner. This work was supported in part by the Swedish Natural Science Research Council.